\documentclass[a4paper,12pt]{article}

\usepackage[english]{babel}
\usepackage[utf8]{inputenc}
\usepackage[T1]{fontenc}
\usepackage{cancel}
\usepackage{multicol}
\usepackage{cite}

\usepackage{float}
\usepackage{authblk}

\usepackage[a4paper,top=2cm,bottom=2cm,left=2cm,right=2cm,marginparwidth=2.cm]{geometry}

\usepackage{setspace}

\usepackage{amsmath}
\usepackage{graphicx}
\usepackage[colorinlistoftodos]{todonotes}
\usepackage[colorlinks=true, allcolors=blue]{hyperref}
\usepackage{lineno}

\title{An improved cosmological parameter inference scheme motivated by deep learning}
\author[1]{Dezső Ribli, Bálint Ármin Pataki, István Csabai}

\begin{document}
\maketitle


\textbf{
Dark matter cannot be observed directly, but its weak gravitational lensing slightly distorts the apparent shapes of background galaxies, making weak lensing one of the most promising probes of cosmology.
Several observational studies have measured the effect, and there are currently running \cite{hildebrandt2016kids,abbott2018dark}, and planned efforts \cite{ivezic2008lsst,laureijs2011euclid} to provide even larger, and higher resolution weak lensing maps.
Due to nonlinearities on small scales, the traditional analysis with two-point statistics does not fully capture all the underlying information \cite{kilbinger2015cosmology}.
Multiple inference methods were proposed to extract more details based on higher order statistics \cite{takada2003three,fu2014cfhtlens},
peak statistics \cite{dietrich2010cosmology, kratochvil2010probing,marian2013cosmological,shan2014weak,liu2015cosmology,kacprzak2016cosmology}, Minkowski functionals \cite{petri2013cosmology, kratochvil2012probing,shirasaki2014statistical} and recently convolutional neural networks (CNN) \cite{schmelzle2017cosmological, gupta2018non}.
Here we present an improved convolutional neural network that gives significantly better estimates of $\Omega_m$ and $\sigma_8$ cosmological parameters from simulated convergence maps than the state of art methods and also is free of systematic bias.
We show that the network exploits information in the gradients around peaks, and with this insight, we construct a new, easy-to-understand, and robust peak counting algorithm based on the 'steepness' of peaks, instead of their heights.
The proposed scheme is even more accurate than the neural network on high-resolution noiseless maps.
With shape noise and lower resolution its relative advantage deteriorates, but it remains more accurate than peak counting.}


\vspace{5mm}

Following the idea and using the simulation data from a recent study \cite{gupta2018non} we created an improved convolutional neural network (CNN) architecture (see details in the Methods) which is able to recover cosmological parameters more accurately from simulated weak lensing maps. 
The input of the network is a set of mock convergence ($\kappa$) maps generated by ray-tracing n-body simulations with 96 different values for the matter density $\Omega_m$ and the scale of the initial perturbations normalized at the late Universe, $\sigma_8$ (see \cite{gupta2018non} and \cite{matilla2016dark} for details of the weak lensing map generation),  the outputs of the network were the predicted cosmological parameters.
The modifications of the CNN mostly consisted of adding further activations, increasing the number of filters, and introducing a regular block structure, following successful computer vision models \cite{krizhevsky2012imagenet, simonyan2014very}.

\begin{figure}[H]
\centering
\includegraphics[width=0.7\textwidth]{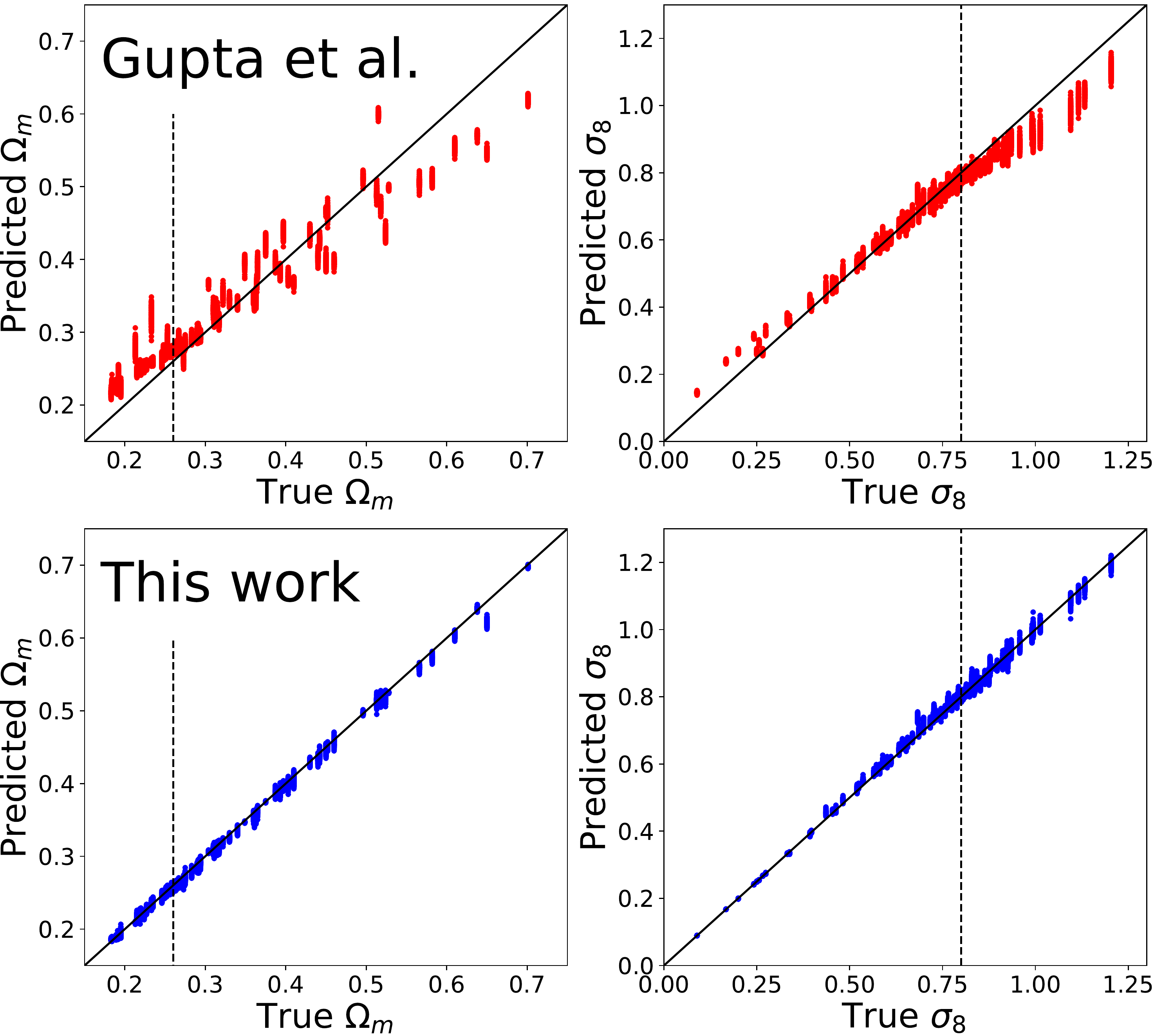}
\caption{\label{fig:improved_cnn} \textbf{Neural network predictions vs. true cosmological parameters used in the simulations.}
\textbf{A}: Predictions made with the neural network from\cite{gupta2018non}.
\textbf{B}:Predictions made with our improved model.
Our CNN shows significantly higher accuracy for both $\Omega_m$ and $\sigma_8$ parameters than the architecture from previous work \cite{gupta2018non}, and it also is free of strong systematic bias.
The dashed lines mark the fiducial parameter values ($\Omega_m=0.26$, $\sigma_8=0.8$) and the solid line represents perfect predictions.}
\end{figure}

Our CNN's parameter estimation accuracy - on previously unseen lensing maps - beats all the state-of-art approaches for the complete parameter range and more importantly, it is free of systematic bias as shown in [Fig. \ref{fig:improved_cnn}] and summarized in [Table \ref{tab:errors}].
The CNN architecture used in the previous work \cite{gupta2018non} shows large errors, and strong bias in the predicted predictions, even close to the fiducial cosmological parameters ($\Omega_m=0.26$, $\sigma_8=0.8$).
The improved CNN architecture predicts both $\Omega_m$ and $\sigma_8$ parameters with $\approx$ $ 2 \times$ smaller errors than peak counting in the full parameter range, with no bias, demonstrating that convolutional neural networks are indeed capable of extracting significantly more information from weak lensing maps than standard approaches.\\


Despite the superb accuracy we demonstrated, the peculiarities of neural networks warrant extraordinary caution when trying to infer credible physical parameters from measurement data with a CNN, which was only trained on simulated data.
%
Interesting results in the context of image recognition caution that neural networks may not be as robust as the regular tool-set of a cosmologist.
It is possible to engineer malicious, imperceptible perturbations of images which completely fool a CNN \cite{goodfellow2014explaining} and it was shown that unexpected inputs to deep neural networks are most likely to be processed incorrectly, and the behavior of a CNN is only reasonable on a thin manifold encompassing the training data \cite{goodfellow2014explaining}.
It is also important to keep in mind that parameter inference with neural networks is conceptually different from the established approaches in weak lensing, which rely on directly comparing simulation data and measurements, through reduction to the power spectrum or peak statistics.
A CNN, on the other hand, learns to approximate a numerical function which directly maps the very high dimensional space of measurements into the final parameter space.
One consequence of this process is that without explicit data comparisons the estimation of the goodness of fit through residuals is not possible.
%
It may be possible to overcome these hurdles through careful investigations, however, we chose to follow a different path in our study.\\


\begin{figure}[H]
\centering
\includegraphics[width=1.0\textwidth]{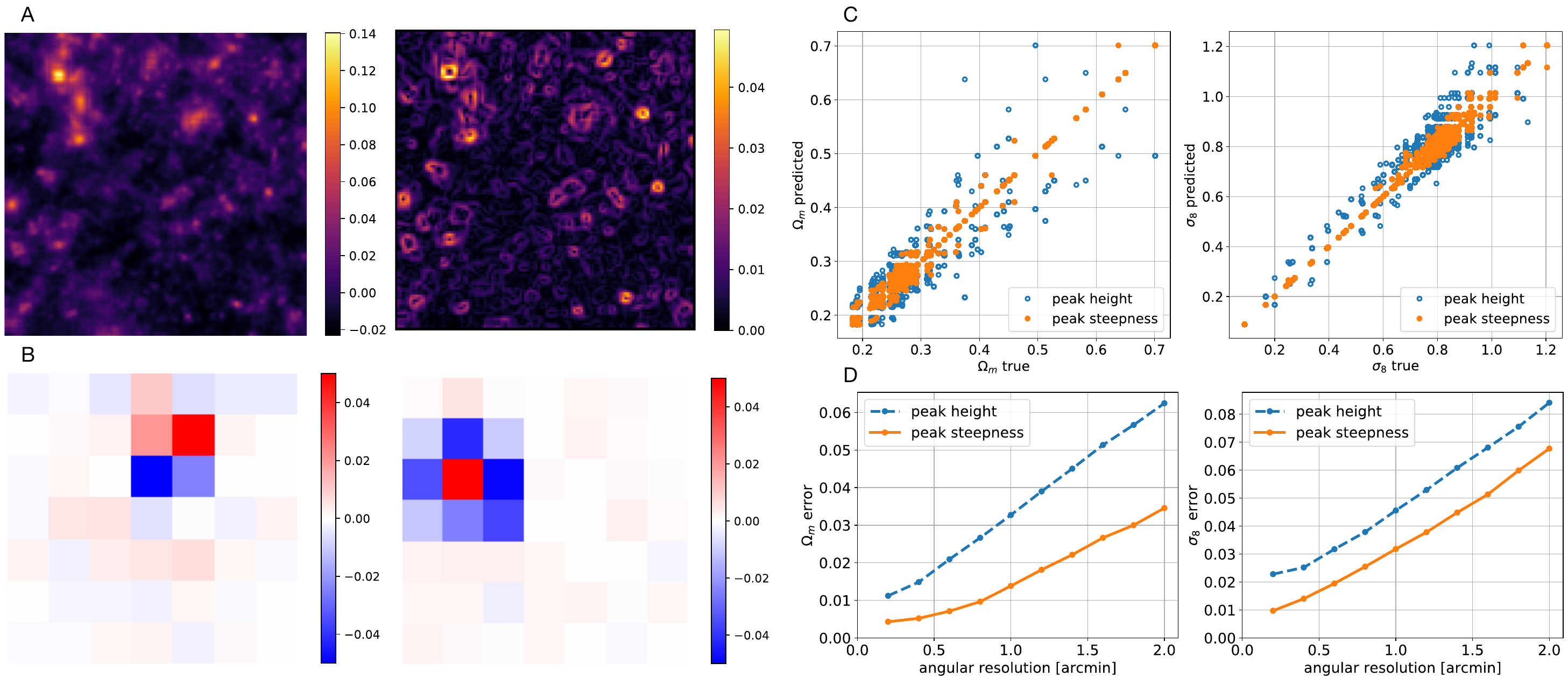}
\caption{\label{fig:noiseless}
\textbf{Depiction of gradient magnitude of the convergence maps, the kernels learned by the CNN, a scatter-plot of predictions with different peak counting schemes, and the effect of angular resolution.}
\textbf{A:} A $128$ pixel sized patch of a noiseless high-resolution convergence map (left), and gradient magnitudes calculated with the Roberts cross kernels on the same map (right).
Peaks are surrounding by circular areas of strong gradients.
\textbf{B:} The two most characteristic kernels in the first layer of the CNN, the (negative) discrete Laplace operator (left), and a Roberts cross kernel (right).
\textbf{C:} Predictions and true cosmological parameter values used in the simulations and with the original and the new peak counting schemes. 
Peak steepness is only shown with the Roberts cross kernel, which has the highest accuracy.
The smaller scatter of the predictions show that the new peak counting scheme has higher accuracy.
The results were calculated on the original 0.2 arcmin resolution, noiseless, simulated convergence maps.
\textbf{D:} The effect of angular resolution on the errors of the different peak counting schemes.
Peak steepness has higher accuracy at lower angular resolutions too.}
\end{figure}

Although the complex interplay of millions of parameters in a CNN simply cannot be fully comprehended, through careful investigation of the internal parameters of the trained neural network, we can gain insights into its workings.
The internal representations of neural networks often have human-understandable interpretations, high-level feature maps frequently learn to selectively detect complex concepts on the images such as legs, wheels or faces \cite{zeiler2014visualizing}. 
We attempted to go even further, not only to find meaningful internal representations but to use them as a hint, and build an easy-to-understand and robust estimation method.
In order to make the interpretation of the CNN's weights easier, we trained a different CNN with a larger  ($7\times7$) kernel size in the first layer, on lensing maps resized to 2 arcmin pixel size, similar to the resolution expected from observations.
The inspection of the kernels immediately revealed that the neural network discovered some interesting and familiar concepts from the training data, [Fig \ref{fig:noiseless}].

The neural network learned to use a kernel strikingly similar to a 2D discrete (negative) Laplace operator, that basically calculates the difference of the peaks and the surrounding pixel values. 
With the middle element scaled to $1$ the learned kernel is the following:

\begin{equation}
K \approx \begin{bmatrix}
-0.05 & -0.25 & -0.06\\
-0.21 & 1       & -0.29  \\
-0.07 & -0.15 & -0.22 \\
\end{bmatrix}
\end{equation}

Using this hint, we considered the most isotropic discrete Laplace operator \cite{lindeberg1990scale}, which is very close to the learned kernel, and a simpler version with zeros in the corners.

\begin{equation}
L_{1} = \frac{-10}{3} \begin{bmatrix}
-0.05 &   -0.2       & -0.05 \\
-0.2 &   1     & -0.2 \\
-0.05 &    -0.2    & -0.05  \\
\end{bmatrix},
L_{2} = -4  \begin{bmatrix}
0 &   -0.25       & 0 \\
-0.25 &   1      & -0.25  \\
0 &    -0.25     & 0  \\
\end{bmatrix} \label{eq:3}
\end{equation}

Another interesting kernel learned by the neural network is very similar to one of the Roberts cross kernels [Fig \ref{fig:noiseless}], which approximate the gradient of an image \cite{roberts1963machine}:

\begin{equation}
R_x= \begin{bmatrix}
0 &   1      \\
-1 &   0      \\
\end{bmatrix},
R_y= \begin{bmatrix}
1 &   0      \\
0 &   -1      \\
\end{bmatrix},
G = \sqrt[]{G_x^2 + G_y^2}, \label{eq:4}
\end{equation}

where the Roberts cross kernels are denoted by $R_{x},R_{y}$, the computed gradients with $G_{x},G_{y}$ and the magnitude of the gradient with $G$.

Kernels \eqref{eq:3}-\eqref{eq:4} calculate the difference of peak values and their surroundings, or the gradients around a peak, therefore they potentially describe the steepness of peaks.
Naturally, the gradients at the peaks are $0$, therefore steepness can be described using the magnitudes of gradient values around peaks.
We found that steepness is strongly correlated with the heights of the peaks, as shown previously\cite{marian2013cosmological},  but it may contain more information than height which is used in the standard peak counting method for parameter estimation in weak lensing.
A previous study has shown that $\Omega_m$ and $\sigma_8$ cosmological parameters have substantial effect on the stacked tangential shear profile of peaks, and their steepness \cite{marian2013cosmological}, and another study showed that using the gradient-moments of the field improve predictions of cosmological parameters \cite{petri2013cosmology}, which may explain why these kernels were discovered and used by the neural network.

Based on these insights we can conclude that the neural network achieves better results than peak counting in part by using representations based on the steepness of the peaks.
Beyond understanding the success of the CNN, the steepness of peaks can be used as a simple and robust descriptor, without the rest of the network and its countless other parameters.
With the kernels \eqref{eq:3}-\eqref{eq:4} learned from the CNN we created a new algorithm that uses the distribution of peak steepness values instead of the peak heights, which is used in the original scheme.

The results achieved with the new peak counting scheme are shown on [Fig \ref{fig:noiseless}] and summarized in Table \ref{tab:errors}.
Using the kernels suggested by the CNN, prediction errors were reduced over 2 fold for both $\Omega_m$ and $\sigma_8$ parameters compared to the original peak counting algorithm. 
Another rather surprising result is that the new peak counting scheme even surpassed the neural network's performance. 
The accurate results indicate that the kernels extracted from the neural network were indeed responsible for its high accuracy, and it was possible to combine the best of both approaches in the new peak counting method.
The gain over the neural network may be explained by the fact that a CNN is not able to explicitly construct histograms of data values, or perform likelihood analysis, which could turn out to be the best approach in this case.

The weak lensing maps used in this study have a very high angular resolution (0.2 arcmin), which is not reachable in experiments due to the low density of observable galaxies \cite{chang2013effective}, therefore we evaluated the reconstruction errors on maps with reduced angular resolutions [Fig \ref{fig:noiseless}].
The new peak counting scheme based on steepness continues to predict $\Omega_m$ and $\sigma_8$ parameters more accurately than peak height at lower angular resolutions too.
The results indicate that the steepness of the lensing map around peaks contains additional information compared to the height of the peaks, even for extended distances.\\

The intrinsic ellipticity of galaxies, 'shape noise', has a profound effect on observations and dominates over the lensing signal.
Since the two filters described previously, Laplace and Roberts cross, were learned from noiseless maps they are not robust to noise, therefore need to engineer a new method to estimate the steepness of peaks in a more noise-resistant way.
Gradient estimation with Sobel filters are known to be robust in the presence of noise, therefore we implemented a peak counting version using these filters.
Similarly to the two Roberts cross kernels the two Sobel filters calculate the gradients in the $x$ and the $y$ direction, and the magnitude of the gradient can be calculated in the same way.
The new method based on Sobel filters is therefore essentially the same as the one based on Robert cross kernel, with the difference that gradients are calculated from a larger area, which leads to more robust estimates, however, it also results in a loss of resolution, therefore Sobel fiters are not optimal in the absence of noise.

\begin{equation}
S_x = \begin{bmatrix}
1 & 0 & -1\\
2 & 0 & -2 \\
1 & 0 & -1 \\
\end{bmatrix},
S_y = \begin{bmatrix}
1 & 2 & 1\\
0 & 0 & 0 \\
-1 & -2 & -1 \\
\end{bmatrix}
\end{equation}

To simulate a realistic measurement, we evaluated predictions with peak heights and Sobel filter based peak steepness on lensing maps corresponding to the fiducial cosmologies with additional shape noise and reduced angular resolution.
For each prediction we averaged the histograms of $37$ individual maps, resulting in an approximately $450$ square-degree simulated footprint, and we measured the RMSE of the estimated parameters from 10000 realizations of noise and randomly selected $37$ maps.

In the first setup, which mimics near future surveys, LSST and EUCLID \cite{ivezic2008lsst,laureijs2011euclid}, we resized the maps to an angular resolution of $1$ arcmins and added Gaussian shape noise at an effective galaxy density $n_{g}=26\, arcmin^{-2}$.
Peak steepness was more accurate than height both when predicting  $\Omega_m$ and  $\sigma_8$ [Table:\ref{tab:errors} ].

Our second setup, which is comparable to currently running observations, KiDS and DES \cite{abbott2017dark,hildebrandt2016kids}, had an angular resolution of $2$ arcmins and a galaxy density $n_{g}=8\, arcmin^{-2}$.
While the advantage of peak steepness deteriorated compared to the first case, it was still more accurate than peak height in term of RMSE both when predicting $\Omega_m$ and  $\sigma_8$ [Table:\ref{tab:errors} ].

\begin{figure}[H]
\centering
\includegraphics[width=0.8\textwidth]{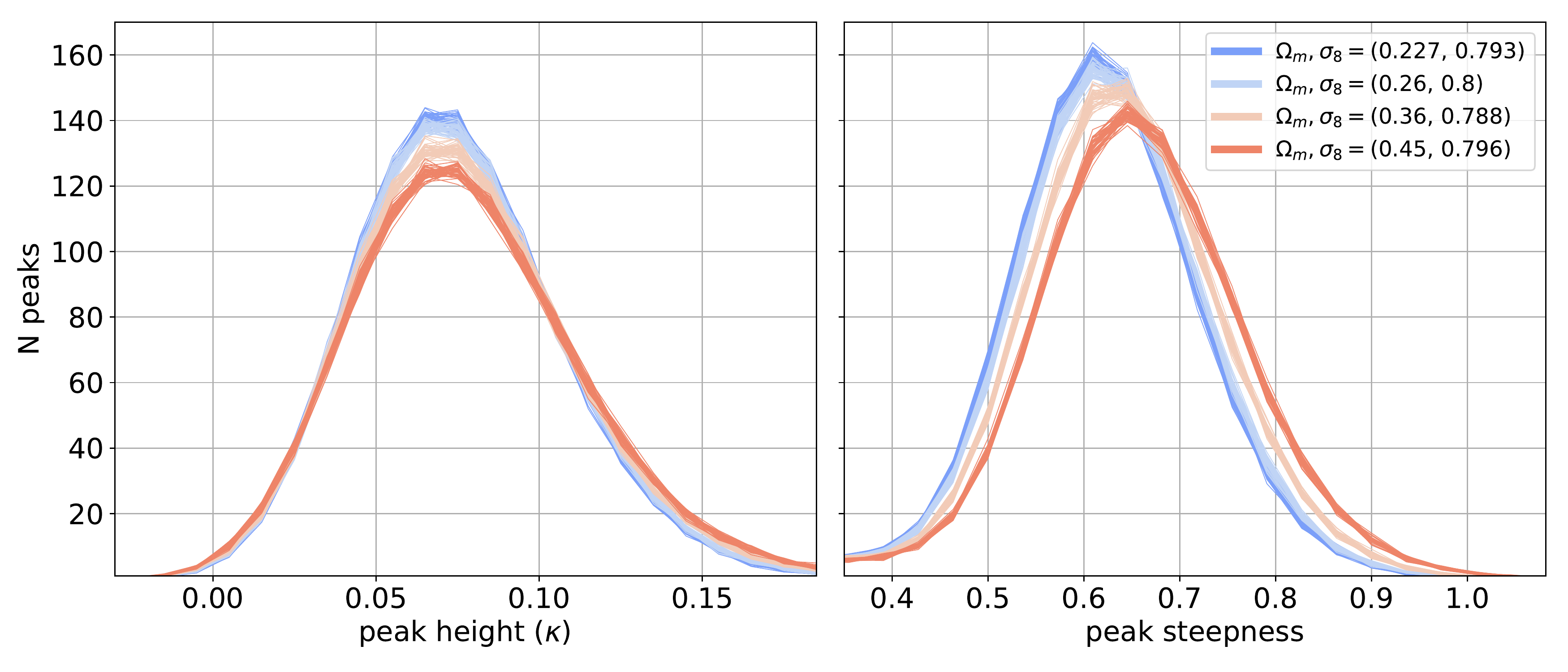}
\includegraphics[width=0.8\textwidth]{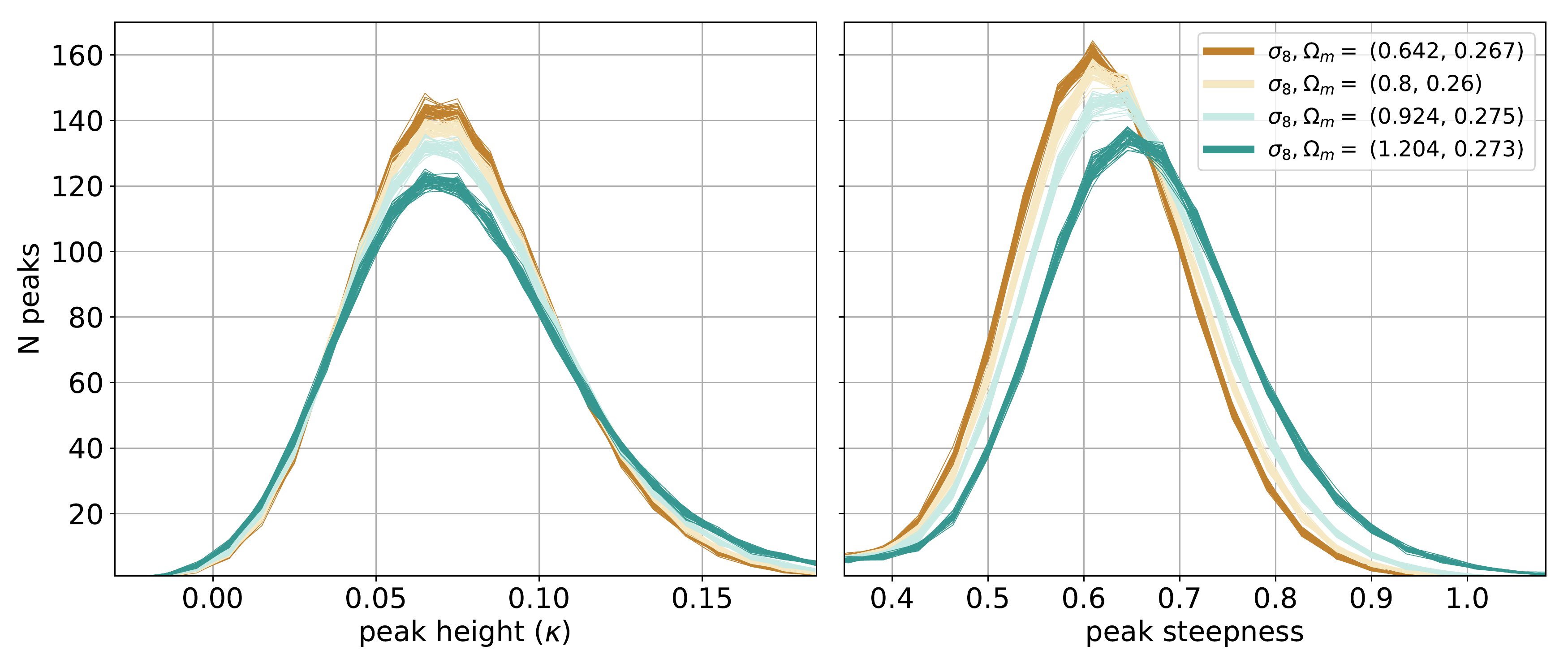}

\caption{\label{fig:noisy}
\textbf{Peak steepness distributions show larger separation between different cosmologies than peak height values, which results in more accurate predictions.}
The distribution of peak steepness values exhibits a shift which is absent in the case of peak heights.  
The peak height (left) and steepness (right) distributions were calculated at 4 different cosmological parameters, with shape noise corresponding to $n_{eff}=8$, at an angular resolution of 2 arcmins.
For each set of parameters 50 noisy realizations calculated from a 450 square degree field are shown.
\textbf{A:} Sensitivity to $\Omega_m$ parameter at similar $\sigma_8$ values.
\textbf{B:} Sensitivity to $\sigma_8$ parameter at similar $\Omega_m$ values
}
\end{figure}

The two scenarios demonstrate that peak steepness with Sobel filtering may be more accurate than peak height even in the presence of shape noise in relevant observational conditions.

\begin{table}[h!]
\centering
\begin{tabular}{l|r|r}
 & $\Omega_m $ (RMSE) $\times 10^3$ & $\sigma_8 $ (RMSE) $\times 10^3$\\\hline
 \hline
Peak counting  & $11.2 \pm 0.2$ & $22.8 \pm 0.2$ \\
CNN \cite{gupta2018non} & $35.1 \pm 0.2$ & $40.3 \pm 0.3$  \\
CNN (ours) & $5.5 \pm 0.1$ & $13.5 \pm 0.1$ \\
Laplace v1  & $4.7 \pm 0.1$ & $11.6  \pm 0.1$  \\
Laplace v2  & $4.6 \pm 0.1$  & $10.9  \pm 0.1$  \\
Roberts cross   & \textbf{4.3} $\pm 0.1$ & \textbf{9.7} $\pm 0.1$ \\\hline
\hline
 & $\Omega_m $ (fiducial) &  $\sigma_8 $ (fiducial) \\\hline
 \hline
Peak counting (noisy, $n_g=8$)  & $43.9 \pm 0.8$ & $64.4\pm 0.7$ \\
Sobel filter  (noisy, $n_g=8$) & $30.3 \pm 0.3$ & $56.7\pm 0.6$ \\\hline
Peak counting (noisy, $n_g=26$)  & $19.6 \pm 0.4$ & $34.5\pm 0.4$ \\
Sobel filter  (noisy, $n_g=26$) & $14.8 \pm 0.2$ & $30.9\pm 0.5$ \\\hline
\hline

\end{tabular}
\caption{\label{tab:errors} 
\textbf{Overview of prediction errors for the CNN and different peak counting schemes.}
The evaluation criterion of predictions is the root mean squared error (RMSE) or predicted values (Methods), standard deviations are estimated from 10000 bootstrap samples.\newline
Results on the noisy maps are calculated at the fiducial cosmological model ($\Omega_m =0.26,  \sigma_8=0.8$), with a simulated footprint of $450\,deg^2$. 
The two pair of noise and angular resolution parameters used are:
( $A_{pix} = 4 \, arcmin^{2}$, $n_{g}=8 \,arcmin^{-2}$)  and ($A_{pix} = 1 \, arcmin^{2}$, $n_{g} = 26 \,arcmin^{-2}$).}
\end{table}

In order to gain insight into the mechanism of the peak steepness counting, we evaluated the mean histograms of the peak heights, and the Sobel filters based steepness values for 4 simulations with different $\Omega_m$, but similar $\sigma_8$ parameters, and 4 simulations with different  $\sigma_8$, but similar  $\Omega_m$ parameters [Fig.\ref{fig:noisy}].
The height of the distribution of steepness values seem to decrease similarly to peak counting with higher $\Omega_m$ values, and in addition, the distribution of peak steepness values significantly shifts towards higher values with at higher $\Omega_m$ parameters.
Peak height and steepness distributions share the $\Omega_m$-$\sigma_8$ degeneracy, and very similar effects appear on the histograms when changing $\sigma_8$ values [Fig.\ref{fig:noisy}].\\

Deep convolutional neural networks are promising new tools for the analysis of 2 or 3-dimensional scientific datasets, and the adopters of this technology need to be aware that small details may create large differences in the quality of predictions.
The complexity of choices makes working with neural networks more like an art with no simple recipe for success.

In many cases it may be hard to obtain truly credible physical parameters with a CNN through direct inference from measurement data, therefore we expect that our approach of building simple and robust descriptors based on the insights gained from interrogating a neural network may be applicable to other scientific machine learning studies.

Peak counting based on the steepness of peaks is significantly more accurate than the one based on the height of peaks on noiseless, high resolution convergence maps.
With shape noise and lower resolution its relative advantage deteriorates but it remains more accurate.
Our results indicate that peak counting based on the steepness of peaks have the potential to tighten the constraints of both $\Omega_m$ and $\sigma_8$ cosmological parameters compared to established methods.
Improved parameter constraints from future surveys could alleviate or strengthen the tension between estimates gained from local and cosmic microwave background measurements 
\cite{maccrann2015cosmic}.
The proposed scheme's efficiency on measurement data needs to be evaluated in future studies.


\bibliographystyle{naturemag}


\textbf{Corresponding author:} Correspondence to Istvan Csabai,  csabai@complex.elte.hu. \\


\textbf{Acknowledgements:} This work was partially supported by National Research, Development and Innovation Office of Hungary via grant OTKA NN 114560 and the National Quantum Technologies Program. The authors thank Z Haiman and JMZ Matilla for making available the simulated weak lensing maps used in this study.  \\


\textbf{Author contributions:} I.C., D.R. and B.A.P. contributed to the conception and design of the study, B.A.P. performed the training and evaluation of neural networks, D.R. conducted the experiments with peak steepness. All authors reviewed the manuscript. \\


\textbf{Author information:}  D.R., B.A.P., I.C., Department of Physics of Complex Systems, Eötvös Loránd University, Budapest.\\

\textbf{Competing interests:}
The authors declare no competing financial interests.


\section*{Methods}

\textbf{Data:} Weak lensing convergence maps were generated with ray tracing from 96 cosmological N-body simulations, with 512 realizations from each simulation \cite{gupta2018non}, these realizations were shown to be quasi-independent \cite{petri2016sample}.
Individual simulations had different pairs of $\Omega_m$ and $\sigma_8$ cosmological parameters.
The parameters were most densely sampled around a 'fiducial model' with ($\Omega_m = 0.26, \sigma_8=0.8$), for more details see \cite{gupta2018non}.
Each simulation had the same initial condition, therefore cosmic variance is neglected, and prediction accuracies may be systematically over-estimated. 
As we considered the comparison of different inference methods, this 
One of the simulations ($\Omega_m = 0.285, \sigma_8=1.134$) was not used as it was an obvious outlier based on \cite{gupta2018non} and our findings too.\\

\textbf{Evaluation:} Predictions were evaluated using the most common evaluation criterion used for regression, the root mean squared error (RMSE) of the predicted values.

\begin{equation*}
RMSE = \sqrt{ \frac{1}{N}  \sum_i \left( y_{i}^{predicted} - y_{i}^{true} \right)^2 } 
\end{equation*}

Where N denotes the number of predictions, and $y$ is the quantity predicted.
The uncertainties of the RMSE values were evaluated with the standard deviation of the RMSE values using 10000 bootstrap samples.
Note that this estimation ignores the variability in the models and therefore it underestimates the uncertainty in the presence of shape noise.\\

\textbf{Training neural networks:}
The lensing maps were split into a training, validation, and test set as 60\%, 10\%, and 30\%. Each 1024 $\times$ 1024 pixel map was tiled into 16 smaller images.
During training, the network handled the tiles individually, and during prediction, the inferred values of the 16 tiles were averaged in order to obtain a final prediction for a whole map.
The networks were trained for 5 epochs with Adam optimizer. The optimizer's parameters were $lr=10^{-4}$, $\beta_1=0.9$, $\beta_2=0.999$. The loss was MSE (mean squared error) for the modified architecture and MAE (mean absolute error) for the architecture from \cite{gupta2018non}, as it was used in their analysis too.\\

\textbf{Improved neural network architecture:} Instead of building a neural network completely from scratch, we decided to use the model from the previous work \cite{gupta2018non} as a starting point, in order to highlight the important differences which radically improve performance.
The architectural guidelines were based on the work of very successful computer vision models \cite{krizhevsky2012imagenet, simonyan2014very}.
\textbf{A,} Firstly we made sure that each convolution is followed by an activation layer.
\textbf{B,} Unusually low number of filters in the first layer might cause under-fitting, therefore we increased the number of filters in the first layer from $4$ to $32$.
\textbf{C,} We introduced a regular block structure, made of subsequent 2 convolutions and a spatial downsampling operation, which is a common structure \cite{krizhevsky2012imagenet, simonyan2014very}.
The 5 convolutional blocks have number of filters (32,64,128,128,128) and the blocks are followed by 3 dense layers with 256, 256 andd 2 filters.
Note that although our model starts with more filters, it has fewer parameters overall (2,649,088  vs 1,467,618), due to less extensive dense layers.

Apart from improvements in the architecture, we also decided to simplify the network, in order to show which features are not absolutely necessary for good performance.
We changed the LeakyReLU operations to simpler ReLU operations.
With a large amount of simulation data available, we also decided to drop the Dropout layers from the network.
Mean absolute error (MAE) loss function was also replaced by the more commonly used mean square error (MSE) cost function.

\textbf{Interpretation of the weights learned by the CNN:} In order to detect strong signals in the first layer of the CNN we trained another neural network, with larger ($7\times7$) kernel size in the first layer instead of the original ($3\times3$) kernel size on noiseless maps resized to 2 arcmin angular resolution.
The larger kernel size allows the inspection of meaningful structures compares to the background levels.
The network was trained with L2 regularization, therefore most weights are decaying to 0 if they are not essential.
We reviewed all the 32 kernel weights of the first layer and selected the ones with the strongest interpretable signals.

\textbf{Peak counting baseline:} We evaluated peak counting statistics on the same dataset similarly to \cite{gupta2018non}. 
Peaks were defined as local maxima on the lensing maps.
For each map, histograms of peak $\kappa$ values were counted in $0.01$ wide bins from $-0.03$ to $0.19$.
Individual histograms were compared to the mean histograms from each simulation, and the parameters corresponding to the simulation with the lowest $\chi$ value were selected as prediction values.
Each realization was evaluated when creating the mean histograms and the covariance matrices.
The $\chi$ values were computed as the following:

\begin{equation}
\chi^2_{\Omega,\sigma} = ( \boldsymbol{y} - \boldsymbol{\mu}_{\Omega,\sigma} ) \, \boldsymbol{C}^{-1}_{\Omega,\sigma} \, ( \boldsymbol{y} - \boldsymbol{\mu}_{\Omega,\sigma} )
\end{equation}

Where $\boldsymbol{y}$ denotes the counts as a vector for a given map, $\boldsymbol{\mu}$ the mean counts for a simulation, and $\boldsymbol{C}$ the covariance matrix of the histograms for a given simulation. 
Note that the covariance matrix is calculated separately for each simulation, instead of using a fixed covariance which is calculated for a selected model as in \cite{gupta2018non}.
We found that varying covariance allows much more precise predictions on noiseless maps, whereas in the presence of noise the varying the difference is reduced, as described previously \cite{matilla2016dark}.
In order to avoid problems when some simulations had no variance in some bins, we used the pseudo-inverse of the covariance matrices instead of their inverse. 
The varying covariance matrix eliminates strong systematic errors when the covariance matrices of different simulations differ significantly.
We verified that using a varying covariance instead of a fixed covariance the prediction errors of the peak counting approach were significantly reduced.\\

\textbf{Peak steepness counting:} The approach is almost the same as peak counting,  the peaks are still located on the original maps, however, we use the histograms of the calculated steepness values, instead of the histograms of the peak height values.
For the discrete Laplace filter, the calculation is straightforward, we convolve the original map with the filter, and take the values at the position of the peaks. 
In the case of the Roberts cross kernel, we apply the $R_x$ and $R_y$ filter on the image and calculate the magnitude of the gradient in the 4 adjacent 2x2 pixel blocks around the peak. For the histograms, we used the sum of the calculated 4 magnitudes.
One small difference compared to peak heights is that calculated values are positive by definition at the local peaks, therefore we adjusted bins to run from 0 to 0.22 for the new values.
Otherwise, the number and width of bins remained the same as for peak counting for the Laplace-filter and the Roberts-cross kernels.
We verified, that a similar bin shift for the original peak counting scheme does not explain the different results, and small differences in the bin width or the range do not noticeably alter the results in any schemes.\\

\textbf{Reducing angular resolution:} When testing on lower angular resolutions, we resized the weak lensing maps to a resolution which is an integer times the original resolution, (1,2,3..).
In this decimating scheme, the new pixel values were calculated as the means of the pixel values in the area corresponding to the new pixel in order to avoid artifacts, like moiré in the low-resolution maps.
Reducing angular resolution this way does not destroy gradients like a Gaussian blur used in \cite{gupta2018non}, therefore the modified peak counting schemes can still work on reduced resolution maps.
We expect that measurement data can be handled in a similar way, which does not destroy differences among neighboring pixels.
Gradient values were observed to become larger at lower resolutions, therefore we increased the bin width linearly from $0.01$ to $0.02$, when changing resolution from $0.02$ to $2.0$ arcmins.\\

\textbf{Peak steepness counting with Sobel filters} Sobel filter caluclate the gradients of an image just as the Roberts cross kernels, therefore the calculations are very similar. 
These kernels use information from a larger area, therefore they produce more robust gradient estimates, making them very useful is the presence of shape noise.
We convolved the image with  $S_x$ and $S_y$ Sobel filters to obtain the gradients, $G_x,G_y$, and calculate the magnitude of the gradient as $G =\sqrt{G_x^2 + G_y^2}$ .
For the histograms we used the mean of the gradient values calculated in the 8 adjacent pixels around peaks.
In case of the 2 arcmin resolution maps, and $n_{eff}=8$ , we used 23  bins with equal width, which run from $0.3$ to $1.1$, and in case of the 1 arcmin resolution maps, and $n_{eff}=26$ , we used 23 bins with equal width, which run from $0.4$ to $1.1$.\\

\textbf{Shape noise:} The noise emerging from the intrinsic ellipticity of galaxies was modeled with a Gaussian noise in pixels \cite{matilla2016dark}.

\begin{equation}
\sigma_{pix} = \frac{\sigma_e }{\sqrt{ 2 A_{pix} n_{g}}}
\end{equation}

where $\sigma_e = 0.4$, $A_{pix}$ is the area of a pixel, and $n_g$ the surface density of galaxies. 
The two pair of parameters used in the study were 
( $A_{pix} = 4 \, arcmin^{2}$, $n_{g}=8 \,arcmin^{-2}$)  and ($A_{pix} = 1 \, arcmin^{2}$, $n_{g} = 26 \,arcmin^{-2}$).

\textbf{Source code:}The source code used in this study is available online at \url{ https://github.com/riblidezso/peak_steepness}.\\


\section*{Data and code availability:}

The data that support the plots within this paper and other findings of this study are available from the corresponding author upon reasonable request.\\

\end{document}